\documentclass[a4paper,10pt]{article}
\usepackage{amsmath}
\DeclareMathOperator{\arccot}{arccot}
\usepackage{graphicx}
\usepackage{amssymb}
\usepackage{amssymb}
\def\be{\begin{eqnarray}}
\def\ee{\end{eqnarray}}

\title{Exactly solvable problems in the momentum space with a minimum uncertainty in position}
\author{M. I. Samar and V. M. Tkachuk\\ Department for Theoretical Physics, \\ Ivan Franko National University of Lviv }
\begin{document}

\maketitle

\begin{abstract}
A new approach in solution of simple quantum mechanical problems in deformed space with minimal length is presented. We propose the generalization of Schro\"edinger equation in momentum representation on the case of deformed Heisenberg algebra with minimal length. Assuming that the kernel of potential energy operator do not change in the case of deformation, we obtain exact solution of eigenproblem of a particle in delta potential as well as double delta potential. Particle in Coulomb like potential is revisited and the problem of inversibility and hermicity of inverse coordinate operator is solved. 


\end{abstract}

\section{Introduction}
Quantum mechanics with modification of the usual canonical commutation relations has attracted a lot of attention recently. 
Such interest is motivated by the investigations in  string theory and quantum gravity,
which suggest the existence of minimal length as a finite lower bound 
to the possible resolution of length \cite{GrossMende,Maggiore,Witten}.
Minimal length can be achieved by modifying usual canonical
commutation relations \cite{Kempf1994,KempfManganoMann,HinrichsenKempf,Kempf1997}.
In present paper we study the simplest deformed algebra 
\be \label{deformation}
 [{X},{P}]=i\hbar(1+\beta{P}^2)
\ee
leading to minimum uncertainty in position $\Delta X_{min}=\hbar\sqrt{\beta}$, which is called minimal length.  

 One can encounter some difficulties connected with the deformation of commutation relations (\ref{deformation})
 while solving the quantum mechanical problems. Therefore, exact solution of the Shro\"dinger equation  have been obtained only for a few systems. 
 They are the one-dimensional harmonic oscillator with minimal uncertainty in position \cite{KempfManganoMann} and also 
 with minimal uncertainty in both position and momentum \cite{Tkachuk2003,Tkachuk2004},
 the D-dimensional isotropic harmonic oscillator \cite{Chang,Dadic}, 
 the three-dimensional Dirac oscillator \cite{Tkachuk2005}, 
 (1+1)-dimensional Dirac oscillator within Lorentz-covariant deformed algebra \cite{Tkachuk2006}, 
 the one dimensional Coulomb problem \cite{Fityo,Pedram} and a particle in the singular inverse square potential \cite{Bouaziz2007,Bouaziz2008}.
 
 Difficulties in obtaining exact solution of quantum mechanical problems with deformed commutation relation forced to develop the perturbation
 techniques and numerical calculus. In \cite{Kempf1997} the perturbational D-dimensional isotropic harmonic oscillator was considered. Three-dimensional Coulomb problem with deformed Heisenberg algebra was studied 
within the perturbation theory in the nonrelativistic case \cite{Brau,Benczik,StetskoTkachuk,Stetsko2006,Stetsko2008} 
and in the case of Lorentz-covariant deformed algebra \cite{SamarTkachuk,Samar}. 
Numerical result for hydrogen atom spectrum in a space with deformed commutation relation was obtained in \cite{Benczik}.

 The studies of quantum mechanical systems in deformed space with minimal length
 is interesting not only from the mathematical point of view, but also from physical one, because they give the possibility to consider the influence of the deformation on the energy spectra. Comparison of  these results with experimental data may yield an estimation of the value of the minimal length(see, for instance, \cite{StetskoTkachuk}).

In present paper we expand the list of exactly solvable quantum mechanical problems in deformed space with minimal length.
In Section II we review Schr\"odinger equation in  momentum representation and generalize it on the case of deformed space.
Next, we consider attractive delta potential and double delta potential in Section III and IV, correspondingly.  In Section IV we revisit one dimensional Coulomb-like problem.
Finally, Section V contains the conclusion.

\section{Schr\"odinger equation in the momentum space}

In ordinary quantum mechanics Schr\"odinger equation can be written in momentum representation as the following integral equation
\be
\frac{p^2}{2m}\phi(p)+\int_{-\infty}^{\infty}U(p-p')\phi(p')dp'=E\phi(p)
\ee
with
\be \label{connection}
U(p-p')=\frac{1}{2\pi\hbar}\int_{-\infty}^{\infty}V(x)\exp\left(-\frac{i}{\hbar}(p-p')x\right)dx
\ee
being the  kernel of the potential energy operator.
In deformed case momentum and position operators can be presented using undeformed ones in the form
\be
&&{P}=\frac{1}{\sqrt{\beta}}\tan\sqrt{\beta}{p},\\
&&{X}={x},
\ee
with $[x,p]=i\hbar$.
We assume the Schr\"odinger equation in the deformed space reads as follows
\be
\frac{1}{2m\beta}\tan^2\left(\sqrt{\beta}p\right)\phi(p)+\int_{-\frac{\pi}{2\sqrt{\beta}}}^{\frac{\pi}{2\sqrt{\beta}}}U(p-p')\phi(p')dp'=E\phi(p).
\ee
Notice, that (\ref{connection}) now is only formally valid, because in deformed space the coordinate representation does not exist due to the minimal length.  
\section{Delta potential in deformed space with minimal length}
In undeformed case attractive  delta potential in the coordinate representation 
\be
V(x)=-2\pi\hbar U_0\delta(x)
\ee
corresponds to constant potential in momentum space due to (\ref{connection})
\be \
U(p-p')=-U_0.
\ee
We assume that in deformed space delta potential is still expressed by constant potential in momentum space. The Schr\"odinger equation then reads
\be \label{Schrodinger_Delta}
\frac{1}{2m\beta}\tan^2\left(\sqrt{\beta}p\right)\phi(p)-U_0\int_{-\frac{\pi}{2\sqrt{\beta}}}^{\frac{\pi}{2\sqrt{\beta}}}\phi(p')dp'=E\phi(p).
\ee
The solution of (\ref{Schrodinger_Delta}) can be easily obtained
\be\label{phi}
\phi(p)=\frac{2m\beta U_0\tilde{\varphi}}{\tan^2\left(\sqrt{\beta}p\right)+\beta q^2}.
\ee
Here we use the notation
\be
\tilde{\varphi}=\int_{-\frac{\pi}{2\sqrt{\beta}}}^{\frac{\pi}{2\sqrt{\beta}}}\phi(p')dp',
\ee
and
\be
q=\sqrt{-2mE}.
\ee
Integrating (\ref{phi}) over $p$ in the range $[-\frac{\pi}{2\sqrt{\beta}},\frac{\pi}{2\sqrt{\beta}}]$, we obtain
\be 
q(1+\sqrt{\beta}q)=2\pi mU_0,
\ee
which yields
\be 
q=\frac{-1+\sqrt{1+8\pi m U_0\sqrt{\beta}}}{2\sqrt{\beta}}.
\ee
Here we omit another root of the quadratic equation because it is negative.

The energy spectrum, like in undeformed case, consist of one energy level
\be E=-\frac{1+4\pi mU_0\sqrt{\beta}-\sqrt{1+8\pi m U_0\sqrt{\beta}}}{4m\beta}. \ee
For small $\beta$ energy spectrum can be approximated as
\be \label{Delta_energy}
E=-2\pi^2mU_0^2+8\pi^3m^2U_0^3\sqrt{\beta}-40\pi^4m^3U_0^4\beta+o(\beta^{3/2}).
\ee
The corresponding normalized wavefunction is
\be\label{normalised phi}
\phi_{}(p)=\sqrt{\frac{2}{\pi}}\frac{\beta(1+\sqrt{\beta}q)}{\sqrt{1+2\sqrt{\beta}q}}\frac{q^{3/2}}{\tan^2\left(\sqrt{\beta}p\right)+\beta q^2}.
\ee
 In the limit $\beta\rightarrow0$ energy $E$ and corresponding wave function $\phi$ go to well known undeformed result.

Note, that this problem was also considered in the linear approximation over the deformation parameter in \cite{Ferkous}. Expression for the energy spectrum obtained in \cite{Ferkous} has other multipliers before $\sqrt{\beta}$ and $\beta$ in comparison with (\ref{Delta_energy}). The reason for this discrepancy of the results is connected with the fact that 
in \cite{Ferkous} $p$ was mistakenly considered to belong to $(-\infty, \infty)$ instead of $(-\frac{\pi}{2\sqrt{\beta}},\frac{\pi}{2\sqrt{\beta}})$.
\section{ Double delta potential}
The problem of two equal, symmetrically displaced  from the origin of coordinate by distance $a$, attractive delta
potentials is described by
\be
V(x)=-\pi\hbar U_0\left(\delta(x-a)+\delta(x+a)\right).
\ee
 This problem corresponds to the harmonic potential in momentum space due to (\ref{connection})
\be \label{harmonic}
U(p-p')=-U_0\cos\left(\alpha (p-p')\right),
\ee 
with $\alpha=a/\hbar$.

We again assume that in deformed space this problem is still expressed by the potential (\ref{harmonic})  in momentum space. The Schr\"odinger equation of the problem can be written as
\be \label{Schrodinger_DoubleDelta}
\frac{1}{2m\beta}\tan^2\left(\sqrt{\beta}p\right)\phi(p)-U_0\int_{-\frac{\pi}{2\sqrt{\beta}}}^{\frac{\pi}{2\sqrt{\beta}}}\cos(\alpha (p-p'))\phi(p')dp'=E\phi(p).
\ee
By changing $p$ for $-p$ in (\ref{Schrodinger_DoubleDelta}), we can show that the Hamiltonian commutes with the parity operator.  Therefore the Hamiltonian and  the parity operator have common system of eigenfunction. Thus  eigenfunctions of considerable problem  can be chosen as even or odd functions.

 Equation (\ref{Schrodinger_DoubleDelta}) can be rewritten as
\be \label{wavefunction_harmonic}
\phi(p)=\frac{m\beta U_0}{\tan^2\sqrt{\beta}p+\beta q^2}\left(\tilde{\varphi}_{-}e^{i\alpha p}+\tilde{\varphi}_{+}e^{-i\alpha p}\right),
\ee
with $q=\sqrt{-2mE}$ and
\be \label{tildevarphi}
\tilde{\varphi}_{\pm}=\int_{-\frac{\pi}{2\sqrt{\beta}}}^{\frac{\pi}{2\sqrt{\beta}}}e^{\pm i\alpha p'}\phi(p')dp'.
\ee
Substituting (\ref{wavefunction_harmonic}) into (\ref{tildevarphi}), we obtain
\be \nonumber
\tilde{\varphi}_{-}=m\beta U_0\left(\tilde{\varphi}_{-}g(0)+\tilde{\varphi}_{+}g(\alpha) \right), \\
\tilde{\varphi}_{+}=m\beta U_0\left(\tilde{\varphi}_{-}g(\alpha)+\tilde{\varphi}_{+}g(0) \right).\label{tildepm}
\ee
Here we use the notation
\be \label{function_g}
g(\alpha)=\int_{-\frac{\pi}{2\sqrt{\beta}}}^{\frac{\pi}{2\sqrt{\beta}}}\frac{\cos{2\alpha p}}{\tan^2\sqrt{\beta}p+\beta q^2}dp.
\ee
Since eigenfunctions of considerable problem  can be chosen as even or odd functions, we demand the following equality for constants $\tilde{\varphi}_{\pm}$ 
\be\label{oddeven} \tilde{\varphi}_{+}=\pm\tilde{\varphi}_{-},\ee 
with upper sign for even and lower one for odd eigenfunctions.
Substituting (\ref{oddeven}) into (\ref{wavefunction_harmonic}) we obtain eigenfunctions of the problem in the form
\be
\phi_{even}(p)=\frac{A}{\tan^2\sqrt{\beta}p+\beta q^2}\cos\alpha p, \\
\phi_{odd}(p)=\frac{B}{\tan^2\sqrt{\beta}p+\beta q^2}\sin\alpha p.
\ee
By substitution of (\ref{oddeven}) into (\ref{tildepm}), we write the equation on the energy spectrum
\be
\frac{1}{\beta mU_0}-g(0)=\pm g(\alpha).
\ee

We failed to obtain analytical expression for integral $g(\alpha)$ in general form. But for some special cases of parameter $\alpha$ we found
\be
g(0)&=&\frac{\pi}{\beta q(1+\sqrt{\beta}q)},\\
g(\sqrt{\beta}n)&=&\frac{\pi(1-\sqrt{\beta}q)^{n-1}}{\beta q(1+\sqrt{\beta}q)^{n+1}},
\ee
with $n=1,2, \dots$.
 
Thus, if the distance between delta wells  is $2a=2n\hbar\sqrt{\beta}$ the equation on the energy spectrum writes
\be \label{condition}
\frac{1}{mU_0}-\frac{\pi}{q(1+\sqrt{\beta}q)}=\pm \frac{\pi(1-\sqrt{\beta}q)^{n-1}}{ q(1+\sqrt{\beta}q)^{n+1}}.
\ee
In the limit of $\beta\rightarrow0$, keeping the distance between delta wells unchanged, the equation on the energy spectrum (\ref{condition}) leads to undeformed one \cite{España}
\be
q=\pi mU_0\left(1\pm e^{-\frac{2qa}{\hbar}}\right).
\ee
Note, that in the limit $a\rightarrow0$ we recover the results
obtained in the previous section for the delta potential.
\section{Coulomb-like problem}
In undeformed space the Schr\"odinger equation of the one dimensional Coulomb-like problem writes
\be
\frac{{p}^2}{2m}\psi(x)-\frac{\alpha}{\hat{x}}\psi(x)=E\psi(x).
\ee
We propose the inverse coordinate operator in the following form
\be\label{1/x}
\frac{1}{x}=v.p.\frac{1}{x}+A\pi\delta(x),
\ee
with A being real constant. 
This definition of the operator $1/x$ correspondes to the following limit
\be
\frac{1}{x}=\lim_{\varepsilon->0}\frac{x+\varepsilon A}{x^2+\varepsilon^2}.
\ee
Note that such proposal ensures hermicity of the operator $1/x$. Also in the case of definition (\ref{1/x}) the following equality is satisfied
\be
\frac{1}{x}x=x\frac{1}{x}=1.
\ee
The kernel of the potential energy operator $V(x)=-\alpha{x}^{-1}$ in momentum representation reads
\be
U(p-p')=-\frac{\alpha}{2\hbar}(2 i \theta(p'-p)-i+A).
\ee
We traditionally assume that this kernel remains unchanged in deformed space and write the Schr\"odinger equation for considerable problem in deformed space with minimal length
\be\label{Coulomb_problem}
\frac{1}{2m\beta}\tan^2\left(\sqrt{\beta}p\right)\phi(p)-\frac{\alpha}{2\hbar}\left[(i+A)\int_{-\frac{\pi}{2\sqrt{\beta}}}^{\frac{\pi}{2\sqrt{\beta}}}\phi(p')dp'-2i\int_{-\frac{\pi}{2\sqrt{\beta}}}^{p}\phi(p')dp'\right]=E\phi(p). \label{eigenequation}
\ee
Differentiating  the latter integral equation we obtain the differential one 
\be
\frac{1}{2m\beta}\left(\tan^2\left(\sqrt{\beta}p\right)\phi(p)\right)'+\frac{i\alpha}{\hbar}\phi(p)=E\phi(p),
\ee
which yields
\be
\phi(p)=\frac{C}{\tan^2(\sqrt{\beta}p)+\beta q^2}\exp\left({-\frac{i\alpha_0}{\sqrt{\beta}(1-\beta q^2)}\left[\frac{1}{\sqrt{\beta} q}\arctan\frac{\tan(\sqrt{\beta}p)}{\sqrt{\beta }q}-\sqrt{\beta}p\right]}\right).\label{eigenfunction_fin}
\ee
Here normalization constant is
\be C=\sqrt{\frac{2}{\pi}}\frac{\beta(1+\sqrt{\beta}q)q^{3/2}}{\sqrt{1+2\sqrt{\beta}q}}.\ee
We also use the notation
 $\alpha_0= \frac{2m\beta\alpha}{\hbar}$ and     $q=\sqrt{-2mE} $.  

Notice, that from equation (\ref{Coulomb_problem}) we can extract the definition of inverse position operator
\be
\frac{1}{X}\phi(p)=-\frac{i}{\hbar}\int_{-\frac{\pi}{2\sqrt{\beta}}}^{p}\phi(p')dp'+c[\phi], 
\ee
with $c[\phi]$ denoting the following functional
\be
c[\phi]=\frac{i+A}{2\hbar}\int_{-\frac{\pi}{2\sqrt{\beta}}}^{\frac{\pi}{2\sqrt{\beta}}}\phi(p')dp'.
\ee
Our definition ensures hermicity of $1/X$ and is in agreement with that proposed in \cite{Fityo}, but here we obtain the explicit form of functional $c[\phi]$, while in \cite{Fityo} this question  was
left outside of consideration. It is interesting that for eigenfunctions (\ref{eigenfunction_fin}) the following equality is true \be
\frac{1}{X}X \phi(p)=X\frac{1}{X}\phi(p)=\phi(p).
\ee
Thus, the problem of inversibility of the  coordinate operator $X$ is solved. 
 
The integrals in  (\ref{eigenequation}) yield
\be
\int_{-\frac{\pi}{2\sqrt{\beta}}}^{\frac{\pi}{2\sqrt{\beta}}}\phi(p')dp'=\frac{2C}{\alpha_0}\sin\varphi_0,
\ee
\be
\int_{-\frac{\pi}{2\sqrt{\beta}}}^{p}\phi(p')dp'=\frac{iC}{\alpha_0}\left(\tan^2\left(\sqrt{\beta}p\right)\phi(p)-e^{i\varphi_0}\right),
\ee
with
$\varphi_0=\frac{\pi\alpha_0}{2\beta q(1+\sqrt{\beta}q)}.$
 Substituting obtained results into equation (\ref{eigenequation}), we find
\be
\sin(\varphi_0-\delta\pi)=0,
\ee
or
\be
\frac{ m\alpha}{\hbar q(1+\sqrt{\beta}q)}=n+\delta, \ \ n=0,1,2...
\ee 
with
\be\delta=\frac{1}{\pi}\arccot{A},\ \ 0 \leq\delta\leq1.\ee
The endpoints  $\delta=0$ and $\delta=1$  are reached in the limit of $A\rightarrow+\infty$ and $A\rightarrow-\infty$, correspondingly.
The energy spectrum is 
\be
E_n=-\frac{1}{8m\beta}\left(1-\sqrt{1+\frac{4m\alpha}{\hbar(n+\delta)}\sqrt{\beta}}\right).
\ee
 Tending $\delta$ to zero for $n= 0$, we obtain that energy goes to $-\infty$ and corresponding eigenfunction vanishes everywhere.  This means that for $\delta=0$ quantum number  $n=1,2, \dots$   and  energy spectra for $\delta=0$  and $\delta=1$  coincides.
Thus, we obtain the families of spectra, which are the same as presented in \cite{Fityo}. Each of these families are characterized
by the value of $\delta$ or equivalently $A$. The values of mentioned parameters
can be calculated from the results of an experiment. In the limit of $\beta\rightarrow0$ we can write for $\psi(x)|_{x=0}$ the following
\be
\psi(x)|_{x=0}=\frac{1}{\sqrt{2\pi\hbar}}\int_{-\infty}^{\infty}\phi(p')dp'=\frac{\sqrt{\hbar} C}{\sqrt{2\pi}mU_0}\frac{1}{\sqrt{1+A^2}}.
\ee
The last formula gives a relation of constant $A$ with  to the value  of eigenfunction  in the origin of  coordinate.
\section{Conclusion}
In this paper we studied the Schr\"odinger equation in momentum representation in deformed space with minimal length. Assuming that the kernel of the potential energy do not change in case of deformed commutation relation, we considered  Dirac delta potential and revisited Coulomb-like potential in deformed space with minimal length.

In the case of the Dirac delta potential we obtained that system has one bound state similarly as in undeformed case. We concluded that this system has an interesting property in deformed space. Namely,  the first correction to
the energy level is proportional to $\sqrt{\beta}$. 
For previously solved problems the correction is proportional to $\beta$ in case of harmonic oscillator, $\beta$ or $\beta\ln\beta$ (for $s$-states) in case of hydrogen atom and $\sqrt{\beta}$ in case of $1D$ potential $1/X$. This fact brings a new possibility to uncover the existence of deformed commutation relations for smaller parameter of deformation.

In case of double delta potential we obtained the wave functions and the equations for energy levels. 
In the limit of parameter of deformation to zero this results coincides with well known undeformed ones.  Tending the distance between delta wells to zero we yield results for the delta potential.

We also proposed the inverse coordinate operator $1/X$ in the form that preserves hermicity and fulfillment of the condition 
\be
\frac{1}{X}X=X\frac{1}{X}=1.
\ee
Our definition of $1/X$ contains one arbitrary real parameter, which means that there exist different extensions of operator $1/X$.
We obtain the same families of spectra for the particle in $1D$ potential $1/X$ as in paper \cite{Fityo}.
Each of them is characterized by a value of free parameter. Different value of this parameter correspond to different extension of operator $1/X$.

\section{Acknowledgement}
We  thank  Dr.  Volodymyr  Pastukhov  for helpful discussion. We also would like to thank to Dr.  Volodymyr  Pastukhov and Hrystyna Gnatenko for careful reading of the manuscript.

\end{document}